\documentclass[12pt]{article}
\usepackage{txfonts}
\usepackage{natbib}
\usepackage{graphicx}
\usepackage{times}
\usepackage{color}
\usepackage{wasysym}
\usepackage{endnotes}
\bibpunct{(}{)}{;}{a}{}{,}
\date{August 13, 2015}
\newcommand{\Lnom}{\hbox{$\mathcal{L}^{\rm N}_{\odot}$}}

\begin{document}

\title{RESOLUTION B2\\ on recommended zero points for the absolute and
  apparent bolometric magnitude scales\\  
{\small\it Proposed by IAU Inter-Division A-G Working Group on Nominal
  Units for Stellar \& Planetary Astronomy}}

\maketitle

%\addtocounter{table}{1}
%\section{INTRODUCTION AND MOTIVATION}

The XXIXth International Astronomical Union General Assembly,

\bigskip

{\bf Noting}

\bigskip

\begin{enumerate}

\item the absence of an exact definition of the zero point for the {\it
  absolute and apparent bolometric magnitude} scales, which has
  resulted in the proliferation of different zero points for
  bolometric magnitudes and bolometric corrections in the literature
  (ranging at approximately the tenth of a magnitude level; see e.g., 
  Bessell, Castelli, \& Plez 1998; Torres 2010),

\item that IAU Commissions 25 and 36 approved identical draft
  resolutions for defining the zero point for the bolometric magnitude
  scale (Andersen 1999), but that the resolution never subsequently
  reached the stage of approval by the IAU General Assembly, and was
  only sporadically adopted within the astronomical community,

\item that recent total solar irradiance measurements have led to a
  revised solar luminosity that differs slightly from the value used
  to set the zero point of the absolute bolometric magnitude scale in the
  Commission 25 and 36 draft resolutions,

\end{enumerate}

\bigskip
\vspace{1.0in}

{\bf Considering}

\bigskip

\begin{enumerate}

\item the need for a standardized absolute and apparent bolometric
  magnitude scale for accurately and repeatably transforming
  photometric measurements into radiative luminosities and
  irradiances, independently of the variable Sun,
  
\item that multiple zero points for bolometric corrections pervade the
  literature due to the lack of a commonly adopted standard zero point 
  for the bolometric magnitude scale,

\end{enumerate}

\smallskip
\bigskip
{\bf Recommends}

\bigskip
\begin{enumerate}

\item to define the zero point of the {\it absolute bolometric
  magnitude} scale by specifying that a radiation source with absolute
  bolometric magnitude\endnote{ ~The notation of $M_{\rm bol}$
    referring to {\it absolute bolometric magnitude} and $m_{\rm bol}$
    referring to {\it apparent bolometric magnitude} was adopted by
    Commission 3 (Notations) at the VIth IAU General Assembly in
    Stockholm in 1938:
    https://www.iau.org/static/resolutions/IAU1938$\_$French.pdf.
    $M_{\rm Bol}$ and $m_{\rm Bol}$ refer specifically to bolometric
    magnitudes defined using the zero points of this resolution.}
  $M_{\rm Bol} = 0$\,mag has a radiative luminosity of exactly

\begin{equation}
L_{\circ}\,=\,3.0128\,\times\,10^{28}\,{\rm W}~.
\end{equation}

and the absolute bolometric magnitude $M_{\rm Bol}$ for a source of
luminosity $L$ (in W) is

\smallskip

\begin{equation}
M_{\rm Bol}\,=\,-2.5\,\log\,(L/L_{\circ})\,=\,-2.5\,\log\,L\,+\,71.197\,425\,...
\end{equation}

\smallskip

The zero point was selected so that the {\it nominal solar
  luminosity}\endnote{ ~Modern spaceborne total solar irradiance (TSI)
  instruments are absolutely calibrated at the 0.03\%\, level (Kopp
  2014). The TIM/SORCE experiment established a lower TSI value than
  previously reported based on the fully characterized TIM instrument
  (Kopp et al. 2005, Kopp \& Lean 2011). This revised TSI scale was
  later confirmed by PREMOS/PICARD, the first spaceborne TSI
  radiometer that was irradiance-calibrated in vacuum at the TSI
  Radiometer Facility (TRF) with SI-traceability prior to launch
  (Schmutz et al. 2013).  The DIARAD/PICARD (Meftah et al. 2014),
  ACRIM3/ACRIMSat (Willson 2014), VIRGO/SoHO, and TCTE/STP-Sat3
  (http://lasp.colorado.edu/home/tcte/) flight instruments are now
  consistent with this new TSI scale within instrument uncertainties,
  with the DIARAD, ACRIM3, and VIRGO having made post-launch
  corrections and the TCTE having been validated on the TRF prior to
  its 2013 launch. The cycle 23 observations with these experiments
  are consistent with a TSI value (rounded to an appropriate number of
  significant digits) and uncertainty of:
  $S_{\odot}$\,=\,1361\,($\pm$\,1)~W\,m$^{-2}$ (2$\sigma$
  uncertainty).  The uncertainty range includes contributions from the
  absolute accuracies of the latest TSI instruments as well as
  uncertainties in assessing a secular trend in TSI over solar cycle
  23 using older measurements.  Combining this total solar irradiance
  value with the IAU 2012 definition of the astronomical unit leads to
  a current best estimate of the mean solar luminosity of
  $L_{\odot}$\,=\,4\,$\pi$\,(1\,au)$^{2}$\,$S_{\odot}$ =
  3.8275\,($\pm$\,0.0014)\,$\times$\,10$^{26}$\,W.  Based on this, a
  {\it nominal solar luminosity} of \Lnom\, =
  3.828\,$\times$\,10$^{26}$\,W is adopted. Using the proposed zero
  point $L_{\circ}$, the nominal solar luminosity $\Lnom$ corresponds
  to bolometric magnitude $M_{\rm
    Bol\,\odot}$\,$\simeq$\,4.739\,996\,...  mag --- i.e.,
  sufficiently close to 4.74 mag for any foreseeable practical
  purpose.}  ($\Lnom = 3.828 \times 10^{26}$\,W) corresponds closely
to absolute bolometric magnitude $M_{\rm Bol\,\odot} = 4.74$\,mag, the
value most commonly adopted in the recent literature (e.g., Bessell,
Castelli, \& Plez 1998; Cox 2000; Torres 2010).

% Support for this value: 
% mbol(Sun) among some references: 
% = 4.77   ; Girardi02/08 
% = 4.7554 ; Mamajek12,Pecaut13
% = 4.75   ; Allen76,Lang92/99,[]Cayrel02,Martins05,Bruntt10
% = 4.74*  ; Durrant81,Bessell98,[]Cox00,VandenBerg03,Cushing05,Masana06,Casagrande06/07/08/10,Torres10
% = 4.72   ; Staizys80,[]Buzzoni10
% = 4.7021 (V=-26.75,Mv=4.8221,BCv=-0.12); Alonso95
% = 4.64   ; Schmidt-Kaler
%
% Hence, since the 1997 definition of Caryel (which
% adopted Mbol = 4.75 for the Sun, there has been little
% use of this value. The majority of studies since then
% seem to be using 4.74. 

\smallskip
  
\item to define the zero point of the {\it apparent bolometric
  magnitude} scale by specifying that $m_{\rm Bol}\,=\,0$\,mag
  corresponds to an {\it irradiance} or {\it heat flux
    density}\endnote{~The terms {\it irradiance} and {\it heat flux
      density} are used interchangeably, both with SI units of
    W\,m$^{-2}$ (Wilkins 1989, Bureau International des Poids et
    Mesures 2006). See also
    https://www.iau.org/publications/proceedings$\_$rules/units/.}  of

\begin{equation}
f_{\circ}\,=\,2.518\,021\,002\,...\,\times\,10^{-8}\,{\rm W\,m^{-2}}
\end{equation}

and hence the apparent bolometric magnitude $m_{\rm Bol}$ for an
irradiance $f$ (in W\,m$^{-2}$) is

% Note: As a check, this zero point is nearly identical to that adopted by:
% Lang (1974) ``Astrophysical Formulae'': 2.52e-8 
% Peacock (1999)  "Cosmological Physics": 2.52e-8  
% Cox (2000) "Allen's Astrophys. Quan.":  2.52e-8 (p.648)
% i.e. within 3 mmag of both. 

\begin{equation}
m_{\rm Bol}\,=\,-2.5\,\log\,(f/f_{\circ})\,=\,-2.5\,\log\,f\,-\,18.997\,351\,...
\end{equation}

The irradiance $f_{\circ}$ corresponds to that measured from an
isotropically emitting radiation source with absolute bolometric
magnitude $M_{\rm Bol} = 0$\,mag (luminosity $L_{\circ}$) at the
standard distance\endnote{~The parsec is defined as exactly
  (648\,000/$\pi$) au (e.g. Cox 2000, Binney \& Tremaine 2008).  Using
  the IAU 2012 Resolution B2 definition of the astronomical unit, the
  parsec corresponds to 3.085\,677\,581\,...\,$\times$\,10$^{16}$\,m.
  As the absolute bolometric magnitude zero point and astronomical
  unit are defined exactly, further digits for the apparent bolometric
  magnitude zero point irradiance $f_{\circ}$ may be calculated if
  needed.}  of 10 parsecs (based on the IAU 2012 definition of the
astronomical unit).

\smallskip

The adopted value of $f_{\circ}$ agrees with some in common use (e.g.,
Lang 1974, Cox 2000) at the level of $<0.1$\%. Using this zero point,
the {\it nominal total solar irradiance} $S_{\odot}^{\rm N}$
(1361\,W\,m$^{-2}$) corresponds to a solar apparent bolometric
magnitude of $m_{\rm Bol\,\odot}\,\simeq\,-26.832$\,mag.\\

\end{enumerate}

\bigskip

\noindent
{\bf References}

\smallskip

\noindent Andersen, J.\ 1999, Transactions of the International
Astronomical Union, Series B, 23, pgs. 141 \& 182

\smallskip

\noindent Bessell, M.\ S., Castelli, F., \& Plez, B.\ 1998, Astronomy
\& Astrophysics, 333, 231

\smallskip

\noindent Binney, J., \& Tremaine, S.\ 2008, Galactic Dynamics: Second
Edition,~ISBN 978-0-691-13026-2 (HB).~Published by Princeton
University Press, Princeton, NJ USA

\smallskip

\noindent Bureau International des Poids et Mesures, 2006, The
International System of Units (SI), 8th edition, Organisation
Intergouvernementale de la Convention du M\'{e}tre

\smallskip

\noindent Cox, A.\,N.\ 2000, Allen's Astrophysical Quantities, 4th
Edition

\smallskip

\noindent Kopp, G. 2014, Journal of Space Weather and Space Climate,
4, A14

\smallskip

\noindent Kopp, G., Lawrence, G., Rottman, G. 2005, Solar Physics,
230, 129

\smallskip

\noindent Kopp, G., \& Lean, J.\ L.\ 2011, Geophysical Research
Letters, 38, L01706

\smallskip

\noindent Lang, K.~R. 1974, Astrophysical Formulae, A Compendium for
the Physicist and Astrophysicist, Springer-Verlag

\smallskip

\noindent Meftah, M., Dewitte, S., Irbah, A., et al.\ 2014, Solar
Physics, 289, 1885

%\noindent Meftah, M., Irbah, A., Hauchecorne, A., et al.\ 2015, Solar
%Physics, 290, 673

\smallskip

\noindent Schmutz W., Fehlmann A., Finsterle W., et al.\ 2013,
AIP Conf. Proc. 1531, p. 624–627, doi:10.1063/1.4804847

\smallskip

\noindent Torres, G. 2010, Astronomical Journal, 140, 1158

\smallskip

\noindent Wilkins, G.~A. 1989, ``The IAU Style Manual (1989): The
Preparation of Astronomical Papers and Reports''

\smallskip

\noindent Willson, R.\ C. 2014, Astrophysics \& Space Science, 352, 341

\theendnotes

\end{document}